\documentclass[runningheads,citeauthoryear]{apinv}
\usepackage{epsfig,cite,graphics}
\usepackage{marvosym} 
\usepackage{amssymb}
\usepackage[utf8]{inputenc}

\begin{document}

\title{On the link between column density distribution \\
and density scaling relation \\in star formation regions}
\titlerunning{Density scaling relation from column density distribution}
\author{Todor Veltchev\inst{1, 3}, Sava Donkov\inst{2} and Orlin Stanchev\inst{1}}
\authorrunning{Veltchev, Donkov \& Stanchev}
\tocauthor{Todor Veltchev, Sava Donkov, Orlin Stanchev} 
\institute{   University of Sofia, Faculty of Physics, 5 James Bourchier Blvd., BG-1164, Sofia
	\and Department of Applied Physics, Technical University, 8 Kliment Ohridski Blvd., BG-1000, Sofia
	\and Zentrum f\"ur Astronomie der Universit\"at Heidelberg, Institute of Theoretical Astrophysics, Albert-\"Uberle-Str. 2, DE-69120, Heidelberg, Germany\newline
	\email{eirene@phys.uni-sofia.bg}  }
\papertype{Submitted on 15.05.2017; Accepted on xx.xx.xxxx}	
\maketitle

\begin{abstract}
We present a method to derive the density scaling relation $\langle n\rangle \propto L^{-\alpha}$ in regions of star formation or in their turbulent vicinities from straightforward binning of the column-density distribution ($N$-pdf). The outcome of the method is studied for three types of $N$-pdf: power law ($7/5\le\alpha\le5/3$), lognormal ($0.7\lesssim\alpha\lesssim1.4$) and combination of lognormals. In the last case, the method of Stanchev et al. (2015) was also applied for comparison and a very weak (or close to zero) correlation was found. We conclude that the considered `binning approach' reflects rather the local morphology of the $N$-pdf with no reference to the physical conditions in a considered region. The rough consistency of the derived slopes with the widely adopted Larson's (1981) value $\alpha\sim1.1$ is suggested to support claims that the density-size relation in molecular clouds is indeed an artifact of the observed $N$-pdf.    
\end{abstract}

\section*{Introduction}
Regions in the interstellar medium with recent star formation (SF) or where star formation might take place have a complex physics. Some key factors to be taken into account for their proper description are: self-gravity, supersonic compressible turbulence, interstellar magnetic fields, thermal and/or external pressure, feedback from newly born stars or supernovae, cosmic rays. All these influence the observed structure of the region: the {\it local} structure, characterized by features like cloudy fragments (clumps, cores), filaments, spurs, as well the {\it general} structure as expressed, e.g., in scaling relations of velocity dispersion, mean density, temperature and other basic quantities. 

An important information on general structure and physics of SF regions and in the molecular clouds (MCs) associated with them could be extracted from analysis of the distribution of gas column density $N$, called usually `probability distribution function' (hereafter, $N$-pdf). It can be derived, in view of the constant gas-to-dust ratio in the interstellar medium (Bohlin et al. 1978), from dust extinction or dust continuum mapping. Such studies in our Galaxy show that the shape of the $N$-pdf is close to lognormal in `quiescent' MCs without recent SF (Kainulainen et al. 2009; Lombardi, Alves \& Lada 2011) or can be fitted through a combination of a lognormal and a power-law (PL) functions in regions with SF activity (Froebrich \& Rowles 2010; Schneider et al. 2013). Theoretical and numerical studies suggest that a lognormal $N$-pdf indicates purely/predominantly turbulent medium (V\'azquez-Semadeni 1994; Federrath et al. 2010) while the development of a PL `tail' is a manifestation of global contraction of the cloud at time-scales, comparable to the free-fall time (Klessen 2000; Ballesteros-Paredes et al. 2011; Kritsuk et al. 2011; Federrath \& Klessen 2013; Girichidis et al. 2014). Thus analysis of the $N$-pdf could provide a clue to the physical conditions in a SF region and/or in its vicinity affected by the interplay of self-gravity and turbulence.

The scaling relation of mean density $\langle n\rangle$ of molecular clouds and cloud fragments, originally discovered by Larson (1981), is considered to be an important indicator of their dynamical state (Myers \& Goodman 1988; Hennebelle \& Falgarone 2012; Kritsuk, Lee \& Norman 2013). It has a power-law form:
\begin{equation}\label{eq_density_scaling_Larson}
 \langle n\rangle \propto L^{-\alpha}~,
\end{equation}
where $L$ is the effective size (scale) and the scaling index $\alpha$ is positive and less or close to unity, within the range $0.1\lesssim L \lesssim10^2$~pc (see Hennebelle \& Falgarone 2012, for review). The value $\alpha\simeq1$ corresponds to ensemble of objects with constant column density $N\sim\langle n\rangle L$. There is an ongoing debate whether such scaling relation is indeed an observational artifact, resulting from use of column-density thresholds to define the clouds or clumps near or abobe the peak of the $N$-pdf (Ballesteros-Paredes \& Mac Low 2002; Lombardi, Alves \& Lada 2010; Ballesteros-Paredes et al. 2012; Schneider et al. 2015). 

Stanchev et al. (2015; hereafter, S15) proposed a method to derive the mean-density scaling relation from analysis of the $N$-pdf in selected zones of Perseus SF region and/or its vicinity. Their approach is novel since it doesn't imply any clump-finding algorithm (i.e. delineation of discrete clumps in the cloud) but introduces abstract spatial scales instead. The procedure includes decomposition of the $N$-pdf to a combination of several lognormal functions of type

\begin{equation}\label{eq_lognormal_component}
{\rm lgn}_{i}(N; a_{i}, N_{i}, \sigma_{i}) = \frac{a_{i}}{\sqrt{2\pi\sigma_{i}^{2}}}\exp{\left(-\frac{[\lg(N/N_{i})]^{2}}{2\sigma_{i}^{2}}\right)}
\end{equation}

where $a_{i}, N_{i}$ and $\sigma_{i}$ are fitting parameters obtained through a $\chi^2$ criterion for goodness. The suggested interpretation has been that each lognormal component (eq. \ref{eq_lognormal_component}) is a signature of spatial domain (scale) with typical column density $N_{i}$. Knowing the effective size $R$ of the studied zone, one can assess the spatial scale attributed to each component

\begin{equation}\label{eq_eff_size_comp_lognormal}
L_{i,\rm \,comp} \equiv L_{\rm comp} (a_i)= \sqrt{\frac{a_{i}}{\sum_{i}{a_{i}}}}R\, 
\end{equation}
and the corresponding mean density 

\begin{equation}\label{eq_density_comp_lognormal}
\langle n \rangle_{\rm comp} \equiv \frac{N_{i}}{L_{\rm comp}(a_i)}~~. 
\end{equation}

The dependence of the S15 outcome on the physical regime and on a possible distance gradient to the studied region were probed in Stanchev et al. (2016). The derived scaling relations in the star-forming region Orion A hint at its gravoturbulent nature.

To sum up, the method of S15 associates spatial scales with discrete components of the column density distribution which account for its local shape (maxima, change of slope). This raises the question whether the proposed approach is {\it physical} or simply {\it morphological}, i.e. reproduces the morphology of the $N$-pdf. If the latter is true, one may device a more straightforward, possibly equivalent technique: to bin the $N$-pdf and define abstract spatial scales $L_{\rm bin}$, proportional to the square root of the total areas which correspond to each bin. Hereafter, we label this approach ($N$-pdf) `binning method'. It comes immediately into question what kind of density scaling relation (if any) yields this method. How would scale the mean density with $L_{\rm bin}$, if defined as

\begin{equation}\label{eq_density_bin}
\langle n \rangle_{\rm bin} \equiv \frac{\langle N\rangle_{\rm bin}}{L_{\rm bin}}~, 
\end{equation}

analogically to the `density of components' $\langle n \rangle_{\rm comp}$ (eq. \ref{eq_density_comp_lognormal})? 

In this work we study the issue in three typical cases of column-density distribution: power law (Section 1), lognormal (Section 2) and a combination of lognormals (Section 3). In the latter case, we test the correlation between the scaling indices of density as obtained from the S15 method (eq. \ref{eq_density_comp_lognormal}) and from the binning method (eq. \ref{eq_density_bin}) and present the result in Section 4. Summary of this work is provided in the Conclusion.

\section*{1. Power-law $N$-pdf}
Such case is idealized, but not far from the real column-density distribution in many regions of recent SF, derived from molecular-line emission. In fact, the actual $N$-pdf shape could be uncertain for densities below the CO self-shielding limit (Lombardi, Alves \& Lada 2015). Then one can consider and analyse the $N$-pdf only in the dynamic range wherein it is power law. 

We define the logarithmic column density $s\equiv \log(N/N_{0})$, adopting some $N_{0}={\rm const}$ within the dynamic range. Its power-law distribution is described by:

\begin{equation}\label{eq_PL_pdf}
dP_{s,\,\rm PL} = A_{s}\left(\frac{N}{N_0}\right)^{q}d\log{\left(\frac{N}{N_0}\right)}=A_{s}\exp(qs)\,ds~,
\end{equation}

where $A_s$ is a normalization coefficient and the slope $q<0$. If the considered range of column densities is divided in logarithmic bins of fixed size $\Delta s$, the total areas of the corresponding domains\footnote{Note that those are not necessarily connected regions.} on the sky are

\begin{equation}\label{eq_volume_bin_PL}
S_{\rm bin}(s)=\int\limits_s^{s+\Delta s}\!\!\!dP_{s^{\prime},\,\rm PL}= \frac{A_s}{q}\,\exp(qs^{\prime}) \Bigg|_s^{s+\Delta s}~,
\end{equation}

with effective sizes (interpreted as abstract scales): 
\begin{equation}\label{eq_eff_size_bin_PL}
L_{\rm bin}(s)=[S_{\rm bin}(s)]^{1/2}= \Bigg[ \frac{A_s}{q}\,\frac{N^q}{N_0^q} \Bigg|_s^{s+\Delta s}~\Bigg]^{1/2}~. 
\end{equation}

Taking {\it small enough} bin size\footnote{Which allows for neglecting of terms of order $(\Delta s)^2$ and higher.} $\Delta s$, one obtains from Taylor-series expansion of the function $N^q(s+\Delta s)$: 

\begin{equation}\label{eq_eff_size-N}
L_{\rm bin}(N)\simeq \Bigg[ \frac{A_s}{q}\,\frac{qN^{q-1}\Delta s}{N_0^q} ~\Bigg]^{1/2}\propto \left(\frac{N}{N_0}\right)^{(q-1)/2}. 
\end{equation}

This leads to a general relation column density vs. size $N_{\rm bin}\propto L_{\rm bin}^{2/(q-1)}$ and -- in view of our definition of mean density (eq. \ref{eq_density_bin}),-- to a density-size relation in the power-law case:
\begin{equation}\label{eq_density-size_PL}
  \langle n \rangle_{\rm bin,\,PL} \propto L_{\rm bin}^{\frac{3-q}{q-1}}~~.
\end{equation}

The extinction distributions, derived from observations of regions with high or moderate SF activity, display pronounced PL parts (Kainulainen et al. 2009). Adopting the standard uniform dust-to-gas ratio (Bohlin et al. 1978), this translates to $N$-pdfs with PL tail at high column densities, typical for the molecular gas phase. Their average slope vary from $q\sim-2$ in clouds associated with H {\sc ii} regions down to $q\gtrsim-4$ for clouds at earlier stages of star formation (Abreu-Vicente et al. 2015; Schneider et al. 2015). Numerical simulations of self-gravitating isothermal and supersonic turbulent clouds also yield -- under assumption of point symmetry, -- an $N$-pdf with PL tail of slope $q=-2.6$ and spanning several orders of magnitude (Kritsuk et al. 2011).  

Thus, from the observational slopes $-2\ge q\ge-4$ in star-forming regions with PL $N$-pdf, one should expect a scaling relation of mean density (eq. \ref{eq_density-size_PL}) with index $7/5\le\alpha_{\rm bin, PL}\le5/3$. Such slopes are substantially steeper than the Larson's one ($\alpha\simeq1$, eq. \ref{eq_density_scaling_Larson}).

\section*{2. Lognormal $N$-pdf}
This case is typical for `quiescent' clouds without star-forming activity (Kainulainen et al. 2009; Schneider et al. 2012), although some lognormal features may appear in the low-density regime due to superposition effects. It points to the turbulent nature of such regions in the interstellar medium as testified by numerical simulations (Federrath et al. 2010). 

The lognormal distribution of column density is given by:

\begin{equation}\label{eq_lognormal_pdf}
dP_{s,\,\rm lgn} = \frac{A_s^\prime}{\sqrt{2\pi\sigma^{2}}}\exp{\left(-\frac{(s-s_{\rm max})^{2}}{2\sigma^{2}}\right)}\,ds~,
\end{equation}
where $A_s^\prime$ is a normalization coefficient, $\sigma$ is the standard deviation and $s_{\rm max}=\log(N_{\rm max}/N_0)$ is the distribution peak.

The total areas of spatial domains, corresponding to bins of size $\Delta s$, are calculated analogically to eq. (\ref{eq_volume_bin_PL}):

\begin{equation}\label{eq_volume_bin_lognormal}
S_{\rm bin}(s)=\int\limits_s^{s+\Delta s}\!\!\!dP_{s^{\prime},\,\rm lgn}= \frac{1}{\pi}\int\limits_t^{t+\Delta t}\!\!\!\exp(-t^{2})\,dt~,~~~t\equiv\frac{s-s_{\rm max}}{\sigma\sqrt{2}}~.
\end{equation}

Hence, the abstract scales in the lognormal case are:
\begin{equation}\label{eq_eff_size_bin_lognormal}
L_{\rm bin}(s)=[S_{\rm bin}(s)]^{1/2}= \Bigg[~\frac{1}{2}\bigg({\rm erfc}(t+\Delta t)-{\rm erfc}(t)\bigg)~\Bigg]^{1/2}~. 
\end{equation}

\begin{figure}[!hb]
  \begin{center}
    \centering{\epsfig{file=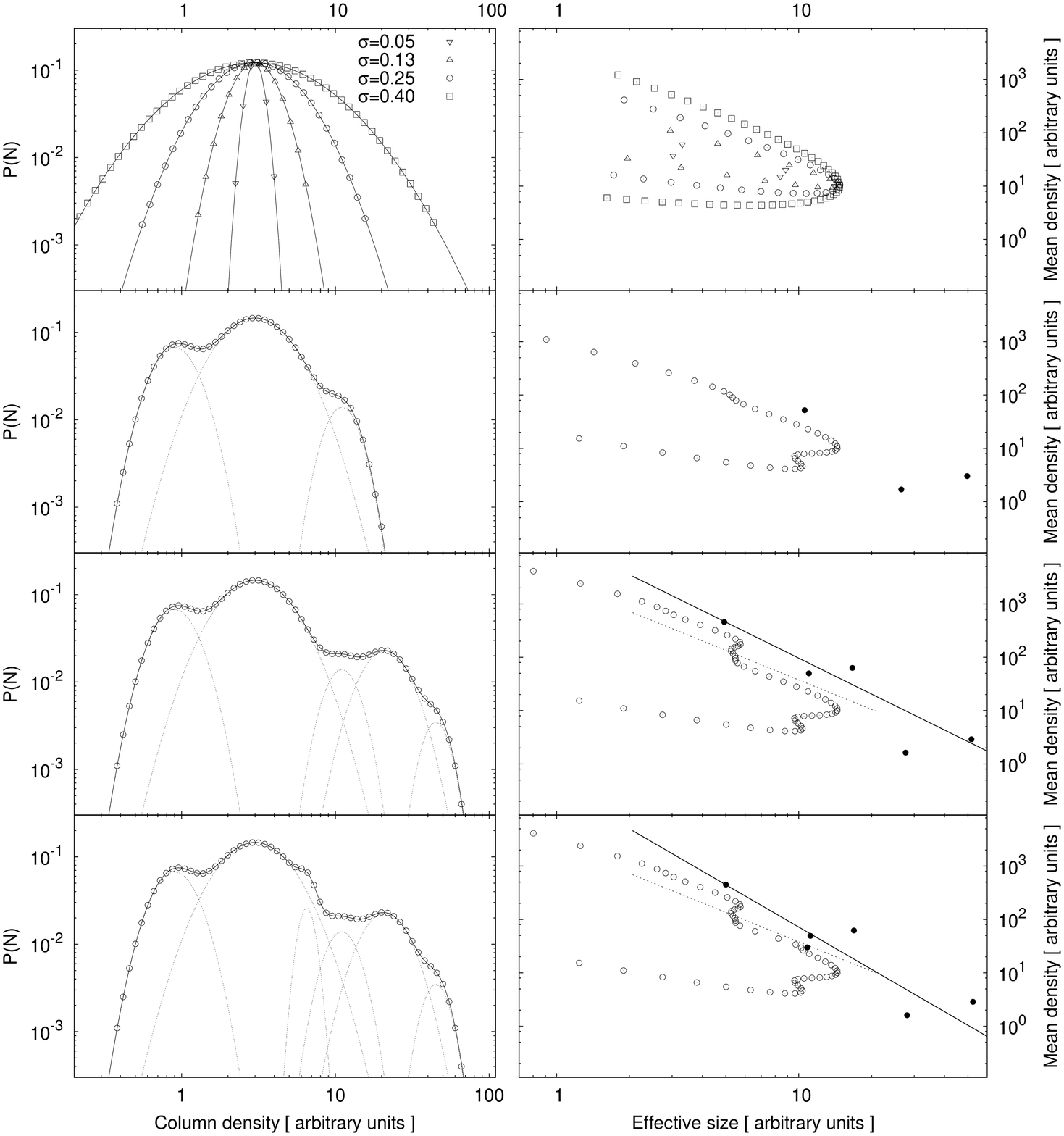, width=1.\textwidth}}
    \vspace{0.4cm}
    \caption[]{Examples of binned $N$-pdfs (left, open symbols), decomposed to one or more lognormal components (dashed; from top to bottom). The corresponding density-size diagrams are displayed in the right column; sizes and densities are calculated from: a) S15 method (eqs. \ref{eq_eff_size_comp_lognormal} and \ref{eq_density_comp_lognormal}; filled symbols); and b) from binning method (open symbols). Weighted power-law fits $\langle n \rangle\propto L^{\alpha}$ are shown for the cases (a) (solid) and (b) (dashed) with 5 and 6 components. See text.}
    \label{fig_density_scaling}
  \end{center}
\end{figure}

In contrast to the PL case (eq. \ref{eq_eff_size_bin_PL}), the scale here is a sophisticated function of $N$ and hence a mean-density scaling relation cannot be obtained analytically. We show in the top row of Fig. \ref{fig_density_scaling} that binning of a single lognormal $N$-pdf (left) results in a double-wing distribution in the density-size diagram (right). The variety of widths $\sigma$ is chosen from observationally derived $N$-pdfs by different authors. Depending on the value of $\sigma$, the scatter of densities is from less than one up to three orders of magnitude and the slope of the density scaling relation varies significantly: $0.7\lesssim\alpha_{\rm bin}\lesssim1.4$. 

\section*{3. $N$-pdf which is a combination of lognormals}
Observational $N$-pdfs\footnote{Or parts of them, excluding the PL tail from consideration.} of regions of more diffuse gas, without signs of active SF can be decomposed to several lognormals (Schneider et al. 2012; Schneider et al. 2013; S15; Stanchev et al. 2016). Typical cases with 3, 5 and 6 components, described by various sets of parameters $a_i$, $\sigma_i$ and $N_i$ (eq. \ref{eq_lognormal_component}), are shown in Fig. \ref{fig_density_scaling}, left (rows 2-4). After binning  of the $N$-pdf, the spatial scales were derived numerically and the corresponding mean densities were calculated from eq. (\ref{eq_density_bin}). The resulting distributions in the density-size diagrams are displayed in Fig. \ref{fig_density_scaling}, right (rows 2-4). With filled symbols we plot the locations of the lognormal $N$-pdf components whose sizes and densities are calculated from eq. (\ref{eq_eff_size_comp_lognormal}) and eq. (\ref{eq_density_comp_lognormal}), respectively. 

If the $N$-pdf is decomposed to 4 components or more, one could compare the slopes of the density scaling relation as derived through the S15 method ($\alpha_{\rm comp}$) and from binning method ($\alpha_{\rm bin}$). To allow for a correct comparison, we introduce weighting of data, proportional to the squared scale size (see Appendix B in S15). The results for exemplary cases with 5 and 6 components (Fig. \ref{fig_density_scaling}, right, rows 3-4) demonstrate that $\alpha_{\rm comp}$ could differ substantially from $\alpha_{\rm bin}$. It is also not clear whether there is a systematic offset between both quantities. This motivated us to perform a test whether $\alpha_{\rm comp}$ and $\alpha_{\rm bin}$ indeed correlate. A negative result of such test would point to a paradigmatic difference between both methods.

\section*{4. Testing the correlation between the methods to derive the density-scaling relation}
\subsection*{4.1. Composing a test sample}
\begin{figure}[!htb]
  \begin{center}
    \centering{\epsfig{file=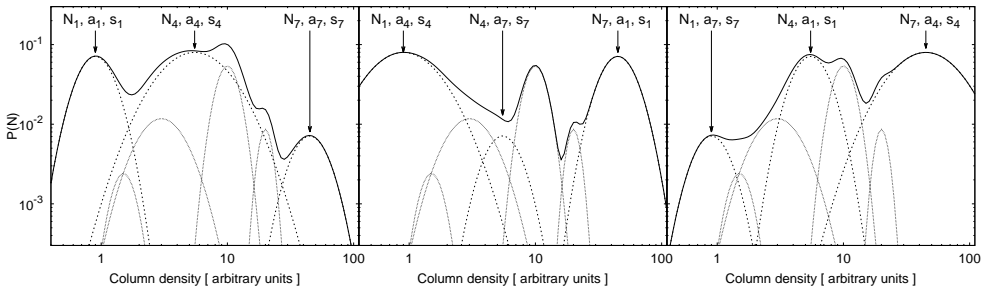, width=1.\textwidth}}
    \vspace{0.4cm}
    \caption[]{Illustration on composing the test sample of $N$-pdfs through permutation of ($a_i,~\sigma_i$) for 3 chosen components (long-dashed). See text.}
    \label{fig_permutation_Ni_test}
  \end{center}
\end{figure}

To achieve reliable result from a correlation test, one needs a statistically significant sample of $N$-pdfs. The latter was composed as follows. 
\begin{itemize}
 \item[$\bullet$] An exemplary $N$-pdf has been chosen which is a combination of 7 lognormal components with parameters $a_i$, $\sigma_i$ and $N_i$ ($1\le i\le7$). This number of components is: i) {\it realistic} for $N$-pdfs in diffuse and/or predominantly turbulent cloud regions (see Appendix in S15), and ii) {\it representative enough} to derive $\alpha_{\rm comp}$ from a fit in the density-size diagram (Fig. \ref{fig_density_scaling}, right; filled symbols).

 \item[$\bullet$] Two further settings regarding the components are: i) the area of the smallest scale to be about 2 orders of magnitude less then the largest, i.e. $(a_i)_{\rm max}/(a_i)_{\rm min}\sim10^2$; ii) the span of column densities be about 2 orders of magnitude, i.e. $(N_i)_{\rm max}/(N_i)_{\rm min}\sim10^2$. Both requirements stem from {\it features of observational $N$-pdfs}; the first one (i) also allows for a reliable detectability of the smallest scales by use of the S15 method. 
 
 \item[$\bullet$] The total areas corresponding to the components have been chosen to constitute a geometric progression, i.e. $a_{i+1}/a_i={\rm const}$ (cf. eq. \ref{eq_eff_size_comp_lognormal}). In that way, we roughly mimic a {\it continuum of scales} within the chosen range.
 
 \item[$\bullet$] Eventually, a sample of $N$-pdfs is produced by permutation of the sets ($a_i,~\sigma_i$), $1\le i\le7$, while keeping the mean column densities $N_i$ of the components fixed. The total number of possible permutations is large: $7!=5040$. We performed only permutations of each 2 and 3 components as shown in Fig. \ref{fig_permutation_Ni_test}. It could be demonstrated that their number is ${{7}\choose{2}}+2{{7}\choose{3}}=91$ which provides the sample size necessary for the correlation test. 

\end{itemize}

\subsection*{4.2. Results}
\begin{figure}[!htb]
  \begin{center}
    \centering{\epsfig{file=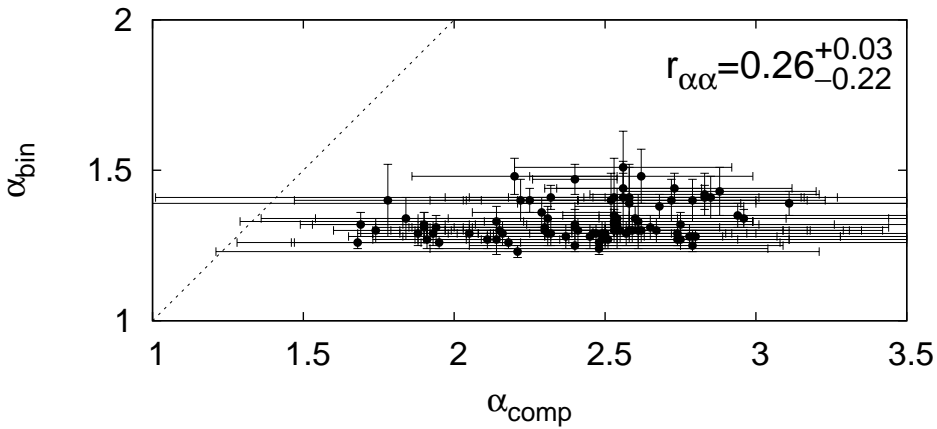, width=1.\textwidth}}
    \caption[]{Comparison between the density scaling indexes $\alpha_{\rm comp}$, derived from the $N$-decomposition to 7 components, and $\alpha_{\rm bin}$ from the $N$-pdf binning. The identity line (dotted) is drawn to guide the eye. See text.}
    \label{fig_alpha-alpha_test}
  \end{center}
\end{figure}

The values of the density-scaling indexes derived for each $N$-pdf from the test sample through the S15 method ($\alpha_{\rm comp}$) and the binning method ($\alpha_{\rm bin}$) are juxtaposed in Fig. \ref{fig_alpha-alpha_test}. While $\alpha_{\rm comp}$ span a large range between about $1.5$ and $3$ (note as well the large uncertainties), the slopes of the density scaling relation from the binning method are constrained within several dex around an average of $1.3$. The correlation is apparently poor. To assess it, we calculated the classical Pearson's coefficient

\begin{equation}\label{eq_Pearson_coefficient}
 r_{\alpha\,\alpha}=\frac{n\sum\limits_{j}(\alpha_{\rm comp})_j(\alpha_{\rm bin})_j-\sum\limits_{j}(\alpha_{\rm comp})_j \sum\limits_{j}(\alpha_{\rm bin})_j}{\sqrt{n\sum\limits_{j}(\alpha_{\rm comp})_j^2-(\sum\limits_{j}(\alpha_{\rm comp})_j)^2}\sqrt{n\sum\limits_{j}(\alpha_{\rm bin})_j^2-(\sum\limits_{j}(\alpha_{\rm bin})_j)^2}}~,
\end{equation}
where the summations are over all $N$-pdfs from the test sample ($n=91$). The uncertainty of $r_{\alpha\,\alpha}$ was estimated by choosing randomly values of $(\alpha_{\rm comp}, \alpha_{\rm bin})_j$ for each $j=1,...,91$ and within the obtained uncertainties of both quantities, for 200 test runs and then adopting the 5nd and 95th percentile of the resulting distribution of $r_{\alpha\,\alpha}$ as lower and upper limits of its value.

The obtained $r_{\alpha\,\alpha}=0.26_{-0.22}^{+0.03}$ does not lead to a clear conclusion from the correlation test. Its value is slightly above the recommended limit ($r_{\alpha\,\alpha}\ge2/\sqrt{n}=0.21$) to accept a significantly non-zero correlation for a sample size $n=91$ (Krehbiel 2004). On the other hand, the high uncertainty with negative sign hints at much lower values of $r_{\alpha\,\alpha}$. Thus a conservative assessment would be that $\alpha_{\rm comp}$ and $\alpha_{\rm bin}$ correlate very week if at all.     

\begin{figure}[!htb]
  \begin{center}
    \centering{\epsfig{file=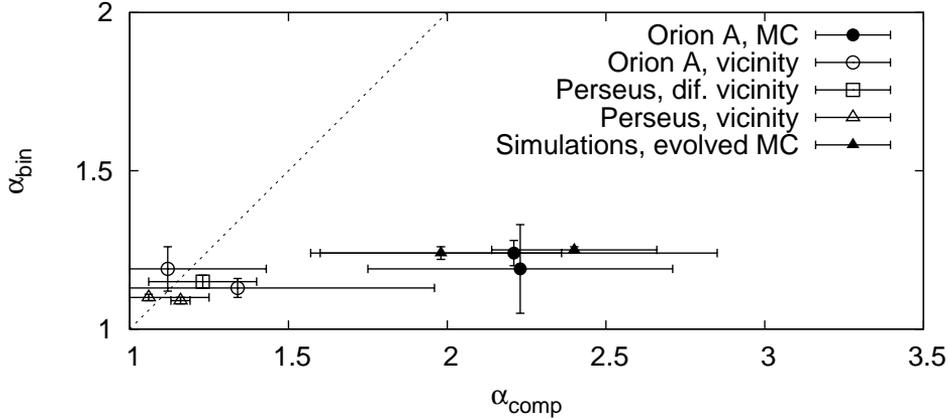, width=1.\textwidth}}
    \caption[]{The same like Fig. \ref{fig_alpha-alpha_test} but from data on the SF region Perseus (Stanchev et al. 2015), molecular cloud (MC) Orion A (Stanchev et al. 2016) and simulations of a evolved molecular cloud (Stanchev et al. 2015). See text.}
    \label{fig_alpha-alpha_obs}
  \end{center}
\end{figure}

In Fig. \ref{fig_alpha-alpha_obs} we compare $\alpha_{\rm comp}$ and $\alpha_{\rm bin}$ from several observed column-density distributions (S15; Stanchev et al. 2016) and from a simulation of evolved molecular cloud with SF (B. K\"{o}rtgen, used in S15). Similar density scaling relations from both methods are obtained for regions of diffuse gas without signs of star formation and possibly with prevalence of turbulent over gravitational energy. Their indexes are close to $\alpha\simeq1.1$ found by Larson (1981) and clearly higher than the average $\alpha\lesssim1$ from many studies which implement clump-finding algorithms (Hennebelle \& Falgarone 2012). A drastic difference between $\alpha_{\rm comp}$ and $\alpha_{\rm bin}$ is obtained for regions with SF activity (Fig. \ref{fig_alpha-alpha_obs}). The indications for a correlation between both methods in regions without SF (open symbols), in view as well of the `quasi-Larson' value of $\alpha$, suggests that the analysis of the $N$-pdf in interstellar medium with prevalent turbulence yields similar results to the ones derived from clump-finding methods (like in Larson 1981 and subsequent studies). This lends support to the idea that the density-size relation in such physical case might be an artifact which reflects features of the $N$-pdf in regions with substantial contribution of low-density zones (Ballesteros-Paredes \& MacLow 2002, Ballesteros-Paredes et al. 2012).
  
\section*{Conclusions}
\label{Conclusion}
This work presents a novel `binning method' for derivation of the density scaling relation $\langle n\rangle \propto L^{-\alpha}$ in regions of star formation and their turbulent vicinities. The technique is based on binning of the column-density distribution ($N$-pdf) as scales and mean densities are defined solely from the local $N$-pdf morphology. It yields a scaling index $\alpha_{\rm bin}$ of the density-size relation as follows:
\begin{enumerate}
 \item In the case of a {\it power-law} $N$-pdf with slope $-2 \le q \le-4$ as found in active star-forming regions, the predicted scaling index $7/5\le\alpha_{\rm bin}\le5/3$ is essentially larger than $\alpha\sim1$ from the seminal work of Larson (1981) and subsequent studies based on clump-finding algorithms.
 \item A {\it lognormal} $N$-pdf with width $\sigma$, indicative for regions without star formation (turbulent clouds), produces shallower slope of the density scaling relation: $0.7\lesssim\alpha_{\rm bin}(\sigma)\lesssim1.4$. We point out that the upper limit is about the lower limit in the power-law case (with $q\sim-4$), which corresponds to an early stage of molecular cloud evolution when gravity slowly takes over. 
 \item In case of a $N$-pdf which can be decomposed to {\it several lognormals}, the index of the density scaling relation varies insignificantly about the upper limit (large widths) in the case of a single lognormal distribution: $1.2\lesssim\alpha_{\rm bin}\lesssim1.5$. 
 \end{enumerate}

In the last case, we perform comparison of the outcome with the one from the method of Stanchev et al. (2015; S15) applied to the same exemplary $N$-pdfs. No clear correlation (if existing at all) is found between the slopes of the density scaling relation derived from the binning method and from the S15 method. The result is interpreted as stemming from the substantially different ideology of the approaches although both are based on analysis of the $N$-pdf. The S15 method is physical and looks for signatures of turbulent scales assuming well developed turbulence and inertial range of scales. In contrast, the `binning method' studied here seems to reflect rather the {\it local morphology} of the column-density distribution with no reference to the physics which shapes the latter. We speculate that the consistency of slopes of the derived relations $\langle n\rangle \propto L^{-\alpha}$ in regions without star formation from both methods with the classical Larson's value $\alpha\sim1.1$ (Larson 1981) from clump-finding techniques supports claims that density-size relation is an artifact of the observed $N$-pdf.

\vspace{6pt}

{\it Acknowledgement:} T.V. acknowledges support by the {\em Deutsche Forschungsgemeinschaft} (DFG) under grant KL 1358/20-1.\\

\end{document}